\documentclass[twocolumn,brazil,english,showkeys, superscriptaddress, nofootinbib]{revtex4}
\usepackage[T1]{fontenc}
\usepackage[latin9]{inputenc}
\usepackage{color}
\usepackage{babel}
\usepackage{mathtools}
\usepackage{amssymb}
\usepackage{graphicx}
\usepackage[unicode=true,pdfusetitle,
 bookmarks=true,bookmarksnumbered=false,bookmarksopen=false,
 breaklinks=true,pdfborder={0 0 0},backref=false,colorlinks=true]
 {hyperref}
\usepackage{breakurl}

\makeatletter
\@ifundefined{textcolor}{}
{%
 \definecolor{BLACK}{gray}{0}
 \definecolor{WHITE}{gray}{1}
 \definecolor{RED}{rgb}{1,0,0}
 \definecolor{GREEN}{rgb}{0,1,0}
 \definecolor{BLUE}{rgb}{0,0,1}
 \definecolor{CYAN}{cmyk}{1,0,0,0}
 \definecolor{MAGENTA}{cmyk}{0,1,0,0}
 \definecolor{YELLOW}{cmyk}{0,0,1,0}
}

\hypersetup{colorlinks=true,citecolor=blue,linkcolor= blue,urlcolor= cyan,filecolor= green}

\makeatother

\begin{document}

\title{Fortran code for generating random probability vectors, unitaries,
and quantum states}

\author{Jonas Maziero}

\email{jonas.maziero@ufsm.br}

\address{Departamento de F\'isica, Centro de Ci\^encias Naturais e Exatas, Universidade Federal de Santa Maria, Avenida Roraima 1000, 97105-900, Santa Maria, RS, Brazil}

\address{Instituto de F\'isica, Facultad de Ingenier\'ia, Universidad de la Rep\'ublica, J. Herrera y Reissig 565, 11300, Montevideo, Uruguay}
\begin{abstract}
The usefulness of generating random configurations is recognized in
many areas of knowledge. Fortran was born for scientific computing
and has been one of the main programming languages in this area since
then. And several ongoing projects targeting towards its betterment
indicate that it will keep this status in the decades to come. In
this article, we describe Fortran codes produced, or organized, for
the generation of the following random objects: numbers, probability
vectors, unitary matrices, and quantum state vectors and density matrices.
Some matrix functions are also included and may be of independent
interest.
\end{abstract}


\maketitle

\section{Introduction}

The generation of random variables has become an essential capability
in fields such as physics, engineering, economics, random search and
optimization, artificial intelligence, and game and network theories
(see e.g. Refs. \cite{Kroese,Hatano_nets,Boguna_nets,Armano_nets,Loffredo_eco,Galam_eco,Rapisarda_eco,Szolnoki_eco,Pham,Rios_opt,Bury_opt,Perc_games,Javarone_games,Silver_games}
and references therein). In Quantum Information Science (QIS), a multidisciplinary
field aiming an efficient and far reaching use and manipulation of
information, the panorama is not different. The creation of random
states and unitaries can be useful for encryption, remote state preparation,
data hiding, classical correlation locking, quantum devices and decoherence
characterization and tailoring, and for quantumness and correlations
quantification \cite{Hayden_RRho,Lloyd_RU,Emerson,Fanchini,Zambrini,Lewenstein,Angelo,Giraud,Wang,Agarwal,Maziero_NMuTP},
to name but a few examples.

Perhaps because of its intuitive syntax and variety of well developed
and optimized tools, Fortran, which stands for Formula translation,
is the primary choice programming language of many scientists. There
are several nice initiatives indicating that it will be continuously
and consistently improved in the future \cite{Szymanski,Metcalf},
what places Fortran as a good option for scientific programming. It
is somewhat surprising thus noticing that Fortran does not appear
in Quantiki's list of ``quantum simulators'' \cite{Quantiki}. For
more details about codes under active development in other programming
languages, see e.g. Refs. \cite{Nori1,Nori2,Gheorghiu,Machnes,Johnston,Fritzsche,Tabakin,Miszczak1,Miszczak2}.
In this article, with the goal of starting the development of a Fortran
Library for QIS, we shall explain (free) Fortran codes produced, or
organized, for generators of random numbers, probability vectors,
unitary matrices, and quantum state vectors and density matrices.
Some examples of free software \cite{FSF} programming languages with
which it would be interesting to develop similar tools are: Python,
Maxima, Octave, C, and Java.

This article is structured as follows. We begin (in Sec. \ref{defs})
recapitulating some concepts and definitions we utilize in the remainder
of the article. In Sec. \ref{sec:general}, the general description
of the code is provided. Reading this section, and the readme file,
would be enough for a black box use of the generators. More detailed
explanations of each one of the them, and of the related options,
are given in Sections \ref{sec:RNG}, \ref{sec:RPVG}, \ref{sec:RUG},
\ref{sec:RSVG}, and \ref{sec:RDMG}. In Sec. \ref{sec:tests} we
summarize the article and comment on some tests for the generators.

\section{Some concepts and definitions}

\label{defs}

In Quantum Mechanics (QM) \cite{Nielsen=000026Chuang,Wilde}, we associate
to a system a Hilbert space $\mathcal{H}$. Every state of that system
corresponds to a unit vector in $\mathcal{H}$. Observables are described
by Hermitian operators $O=\sum_{j}o_{j}|o_{j}\rangle\langle o_{j}|$,
i.e., $o_{j}\in\mathbb{R}$ and $|o_{j}\rangle$ form an orthonormal
basis. Born's rule bridges theory and experiment stating that if the
system is prepared in the state $|\psi\rangle=\sum_{j}c_{j}|o_{j}\rangle$
and $O$ is measured, then the probability for the outcome $o_{j}$
is $p_{j}=|c_{j}|^{2}=|\langle o_{j}|\psi\rangle|^{2}$. We recall
that a set of numbers $p_{j}$ is regarded as a discrete probability
distribution if all the numbers $p_{j}$ in the set are non-negative
(i.e., $p_{j}\ge0$) and if they sum up to one (i.e., $\sum_{j}p_{j}=1$).
In QM, preparations and tests involving incompatible observables lead
to quantum coherence and uncertainty and to the consequent necessity
for the use of probabilities.

When we lack information about a system preparation, a complex positive
semidefinite matrix ($\rho\ge0)$ with unit trace $\mathrm{Tr}(\rho)=1$,
dubbed the density matrix, is the mathematical object used to describe
its state \cite{Nielsen=000026Chuang,Wilde}. In these cases, if the
pure state $|\psi_{j}\rangle$ is prepared with probability $p_{j}$,
all measurement probabilities can be computed in a succinct way using
the density operator $\rho=\sum_{j}p_{j}|\psi_{j}\rangle\langle\psi_{j}|.$
The ensemble $\{p_{j},|\psi_{j}\rangle\}$ leading to a given $\rho$
isn't unique. But, as $\rho$ is an Hermitian matrix, we can write
its unique eigen-decomposition $\rho=\sum_{j=1}^{d}r_{j}|r_{j}\rangle\langle r_{j}|$
with $r_{j}$ being a probability distribution and $|r_{j}\rangle$
an orthonormal basis. We observe that the set of vectors with properties
equivalent to those of $(r_{1},\cdots,r_{d})$, which are dubbed here
probability vectors, define the unit simplex. 

The mixedness of the state of a system follows also when it is part
of a bigger-correlated system. Let us assume that a bi-partite system
was prepared in the state $|\psi_{ab}\rangle$. All the probabilities
of measurements on the system $a$ can be computed using the (reduced)
density matrix obtained taking the partial trace over system $b$
\cite{Maziero_ptr}: $\rho^{a}=\mathrm{Tr}_{b}(|\psi_{ab}\rangle\langle\psi_{ab}|).$

Up to now, we have discussed some of the main concepts of the kinematics
of QM. For our purposes here, it will be sufficient to consider the
quantum mechanical closed-system dynamics, which is described by a
unitary transformation \cite{Nielsen=000026Chuang,Wilde}. If the
system is prepared in state $|\psi\rangle$, its evolved state shall
be given by: $|\psi_{t}\rangle=U|\psi\rangle,$ with $UU^{\dagger}=\mathbb{I}$,
where $\mathbb{I}$ is the identity operator in $\mathcal{H}$. The
unitary matrix $U$ is obtained from the Schr\"odinger equation $i\hbar\partial U/\partial t=HU$,
with $H$ being the system Hamiltonian at time $t$. Between preparation
and measurement (reading of the final result), a Quantum Computation
(in the circuit model) is nothing but a unitary evolution; which is
tailored to implement a certain algorithm.

\section{General description of the code}

\label{sec:general}

The code is divided in five main functionalities, which are: the random
number generator (RNG), the random probability vector generator (RPVG),
the random unitary generator (RUG), the random state vector generator
(RSVG), and the random density matrix generator (RDMG). Below we describe
in more details each one of these generators, and the related available
options. 

A module named \emph{meths} is used in all calling subroutines for
these generators in order to share your choices for the method to
be used for each task. A short description of the methods and the
corresponding options, \texttt{opt\_rxg} (with \texttt{x} being \texttt{n},
\texttt{pv}, \texttt{u}, \texttt{sv}, or \texttt{dm}), is included
in that module. To call any one of these generators, include \texttt{call
rxg(d,rx)} in your program, where \texttt{d} is the dimension of the
vector or square matrix \texttt{rx}, which is returned by the generator.
If you want, for example, a random density matrix generated using
a ``standard method'' just \texttt{call rdmg(d,rdm)}; the same holds
for the other objects. If, on the other hand, you want to choose which
method is to be used in the generation of any one of these random
variables, add \texttt{use meths} after your (sub)program heading,
declare \texttt{opt\_rxg} as \texttt{character(10)}, and add \texttt{opt\_rxg
= \textquotedbl{}your\_choice\textquotedbl{}} to your program executable
statement section.

\section{Random number generator}

\label{sec:RNG}

Beforehand we `have' to initialize the RNG with \texttt{call rng\_init()};
remember to do that also after changing the RNG. As \texttt{rn} is
an one-dimensional double precision array, if you want only one random
number (RN), then just set $d=1$. As the \emph{standard} pseudo-random
number generator (pRNG), we use the Fortran implementation by Jose
Rui Faustino de Sousa of the Mersenne Twister algorithm introduced
in Ref. \cite{Matsumoto}. This pRNG has been adopted in several software
systems and is highly recommended for scientific computations \cite{Katzgraber_RNGs}.
As less hardware demanding alternatives, we have also included the
Gnu's standard pRNG KISS \cite{Marsaglia} and the Petersen's Lagged
Fibonacci pRNG \cite{Petersen}, which is available on Netlib. The
options \texttt{opt\_rng} for these three pRNGs are, respectively,
\texttt{\textquotedbl{}mt\textquotedbl{}}, \texttt{\textquotedbl{}gnu\textquotedbl{}},
and \texttt{\textquotedbl{}netlib}\textquotedbl{}. The components
of \texttt{rn} provided by these pRNG are uniformly distributed in
$[0,1]$. Because of their use in the other generators, we have also
implemented the subroutines \texttt{rng\_unif(d,a,b,rn)}, \texttt{rng\_gauss(d,rn)},
\texttt{rng\_exp(d,rn)}, which return $d$-dimensional vectors of
random numbers with independent components possessing, respectively,
uniform in $[a,b]$, Gaussian (standard normal), and exponential probability
distributions (see examples in Fig. \ref{fig:rn_rpv}).

\section{Random probability vector generator}

\label{sec:RPVG}

Once selected the RNG, it can be utilized, for instance, for the sake
of sampling uniformly from the unit simplex. That is to say, we want
to generate random probability vectors (RPV)
\begin{equation}
\vec{p}=(p_{1},\cdots,p_{d})
\end{equation}
with $p_{j}\ge0$ and $\sum_{j=1}^{d}p_{j}=1$; and the picked points
$\vec{p}$ should have uniform density in the unit simplex. In the
following, we describe briefly some methods that may be employed to
accomplish (approximately) this task.

Let us start with a trigonometric approach to create RPVs (\texttt{opt\_rpvg
= \textquotedbl{}trig\textquotedbl{}}). First we get the angles $\theta_{0}=0$
and $\theta_{j}=\arccos\sqrt{r_{j}}$ (for $j=1,\cdots,d-1$), with
$r_{j}$ being uniform RNs in $[0,1]$. Then we define the components
of the RPV: $p_{j}=\sin^{2}\theta_{j-1}\Pi_{k=j}^{d-1}\cos^{2}\theta_{k}$
(for $j=1,\cdots,d-1$) and $p_{d}=\sin^{2}\theta_{d-1}$ \cite{Vedral_trig}.
To get rid from the bias existing in the generated RPVs we use a random
permutation of $\{1,2,\cdots,d\}$ to shuffle its components \cite{Maziero_rpv_bjp}. 

The normalization method (\texttt{opt\_rpvg = \textquotedbl{}norm}\textquotedbl{})
starts from the defining properties of a probability vector and uses
the RNG to draw uniformly $p_{1}\in[0,1]$, $p_{j}\in[0,1-\sum_{k=1}^{j-1}p_{k}${]}
(for $j=1,\cdots,d-1$), and set $p_{d}=1-\sum_{k=1}^{d-1}p_{k}$.
At last we use shuffling of the components of $\vec{p}$ to obtain
an unbiased RPV \cite{Maziero_rpv_bjp}. A somewhat related method,
which is used here as the \emph{standard} one for the RPVG, was proposed
by \.{Z}yczkowski, Horodecki, Sanpera, and Lewenstein (ZHSL) in the
Appendix A of Ref. \cite{Zyczkowski_vol}; so\texttt{ opt\_rpvg =
\textquotedbl{}zhsl\textquotedbl{}}. The basic idea is to consider
the volume $\Pi_{j=1}^{d-1}\mathrm{d}(p_{j}^{d-j})$ and $d-1$ uniform
random numbers $r_{j}$ and to define $p_{1}=1-r_{1}^{1/(d-1)}$ and
$p_{j}=(1-r_{j}^{1/(d-j)})(1-\sum_{k=1}^{j-1}p_{k})$ (for $j=2,\cdots,d-1$),
and finally making $p_{d}=1-\sum_{k=1}^{d-1}p_{k}$.

\begin{figure}
\includegraphics[scale=0.41]{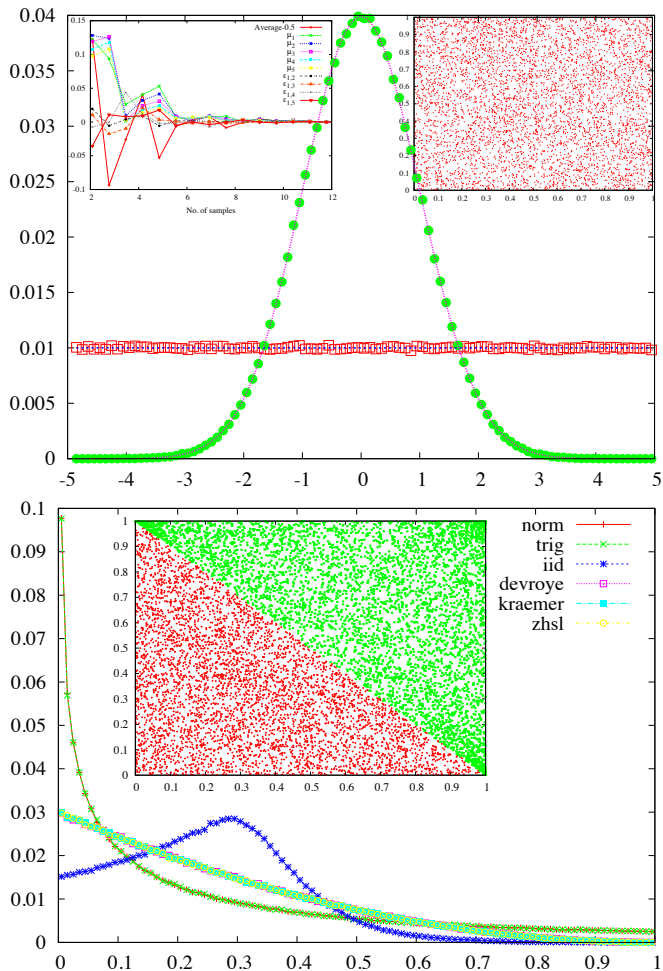}

\caption{\textbf{On top}: Gaussian and uniform probability densities for one
million random numbers generated using the Mersenne-Twister random
number generator. In the right \emph{inset} is shown the 2D scatter
plot for 5000 pairs of RNs generated via the Gnu's RNG. The left inset
shows the shifted mean and some moments, $\mu_{j}$, and correlations,
$\varepsilon_{j,k}$, as a function of the number of samples obtained
with the Netlib RNG. \textbf{On bottom}: Probability density for the
first component of one million random probability vectors with dimension
$d=4$ and generated using the method indicated in the figure (refer
to the text for more details). In the \emph{inset} is shown the 2D
scatter plot for the first two components $(p_{1},p_{2})$ of five
thousand RPVs with $d=3$ and produced using the ZHSL (red) or the
Normalization (green) method (in the last case the points are $(1-p_{1},1-p_{2})$). }

\label{fig:rn_rpv}
\end{figure}

Other possible approach is taking $d$ independent and identically
distributed uniform random numbers $r_{j}$ (thus\texttt{ opt\_rpvg
= \textquotedbl{}iid\textquotedbl{}}) and just normalizing the distribution,
i.e., $p_{j}:=r_{j}/(\sum_{k=1}^{d}r_{k})$ \cite{Maziero_rpv_bjp}.
A related sampling method was put forward in Ref. \cite{Devroye}
by Devroye (\texttt{opt\_rpvg = \textquotedbl{}devroye\textquotedbl{}});
see also the Appendix B of Ref. \cite{Shang}. The procedure is similar
to the previous one, but with the change that the random numbers $r_{j}$
are drawn with an exponential probability density. Yet another way
to create a RPV, due to Kraemer (\texttt{opt\_rpvg = \textquotedbl{}kraemer\textquotedbl{}})
\cite{Kraemer} (see also Refs. \cite{Smith,Grimme}), is to take
$d-1$ random numbers uniformly distributed in $[0,1]$, sort them
in nondecreasing order, use $r_{0}=0$ and $r_{d}=1$, and then defining
$p_{j}=r_{j}-r_{j-1}$ for $j=1,\cdots,d$. For sorting we adapted
an implementation of the Quicksort algorithm from the Rosetta Code
Project \cite{RosettaCode}.

With exception of the iid, all these methods lead to fairly good samples.
With regard to the similarity of the probability distributions for
the components of the RPVs generated, one can separate the methods
in two groups: (a) ZHSL, Kraemer, and Devroye, and (b) trigonometric
and normalization. Concerning the choice of the method, it is worth
mentioning that for moderately large dimensions of the RPV, the group
(a) excludes the possibility of values of $p_{j}$ close to one. This
effect, which may have unwanted consequences for random quantum states
generation, is less pronounced for the methods (b), although here
the problem is the appearance of a high concentration of points around
the corners $p_{j}=0$ (see Fig. \ref{fig:rn_rpv}).

If $\mathcal{R}(N)$ is the computational complexity (CC) to generate
$N$ RNs and $\mathcal{O}(N)$ is the CC for $N$ scalar additions,
then for $d\gg1$ we have the following estimative: $\mbox{CC}(\mbox{RPVG})\approx\mathcal{R}(d)+\mathcal{O}(d\log d)$.

\section{Random unitary generator}

\label{sec:RUG}

A complex matrix $U$ is unitary, i.e.,
\begin{equation}
U^{\dagger}U=\mathbb{I},
\end{equation}
with $\mathbb{I}$ being the identity matrix, if and only if its column
vectors form an orthonormal basis. So, starting with a complex matrix
possessing independent random elements which have identical Gaussian
(standard normal) probability distributions, we can obtain a random
unitary matrix (RU) via the QR factorization (QRF) \cite{Cybenko,Mezzadri}.
We implemented it using the modified Gram-Schmidt orthogonalization
(\texttt{opt\_rug = \textquotedbl{}gso\textquotedbl{}}) \cite{Diaconis,Golub},
which is our \emph{standard} method for generating random unitaries.
We also utilized LAPACK's implementation of the QRF via Householder
reflections (\texttt{opt\_rug = \textquotedbl{}hhr\textquotedbl{}});
so you will need to have LAPACK installed \cite{LAPACK}. Random unitaries
can be obtained also from a parametrization for $U(d)$. We have implemented
a RUG in this way using the Hurwitz parametrization (\texttt{opt\_rug
= \textquotedbl{}hurwitz\textquotedbl{}}); for details see Refs. \cite{Zyczkowski_RU,Zyczkowski_RU_int}.
Here a rough estimate for the computational complexity is: $\mbox{CC}(\mbox{RUG})\approx\mathcal{R}(d^{2})+\mathcal{O}(d^{4})$. 

\begin{figure}
\includegraphics[scale=0.41]{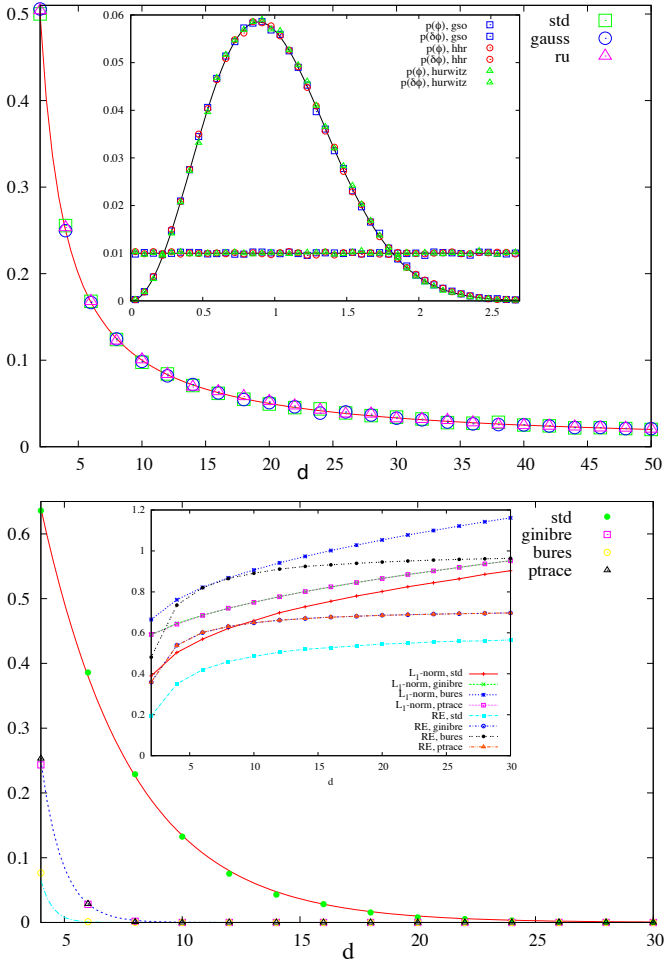}

\caption{\textbf{On top}: Average fidelity, $\langle F(|\psi\rangle,|\phi\rangle)\rangle=\langle|\langle\psi|\phi\rangle|^{2}\rangle$,
as a function of the dimension $d$ for one thousand pairs of random
state vectors generated using the indicated method. The continuous
line is for $1/d$. In the \emph{inset} is shown the probability density
for the eigen-phases and its spacings (divided by the average) for
ten thousand $20\mathrm{x}20$ random unitary matrices. \textbf{On
bottom}: Probability of finding a positive partial transpose bipartite
state of dimension $d=d_{a}d_{b}$, with $d_{a}=2$, for ten thousand
random density matrices produced for each value of $d$. The continuous
lines are the exponential fits, $p=\alpha\mathrm{e}^{-\beta d}$,
with $(\alpha,\beta)$ being $(1.81,0.26)$, $(18.77,1.08)$, and
$(265.21,2.08)$ for, respectively, the \texttt{std}, \texttt{ginibre}
(\texttt{ptrace}), and \texttt{bures} method. In the \emph{inset}
is shown the average $L_{1}$-norm quantum coherence $C_{l1}(\rho)=\sum_{j\protect\ne k}|\rho_{j,k}|$
(divided by $\log_{2}d$) and the relative entropy of quantum coherence
$C_{re}(\rho)=S(\rho_{diag})-S(\rho)$, with $S(\rho)=-\mathrm{Tr}(\rho\log_{2}\rho)$
being von Neumann's entropy and $\rho_{diag}$ is obtained from $\rho$
by erasing its off-diagonal matrix elements, in basis $|j\rangle$
($10^{4}$ samples were produced for each value of $d$).}

\label{fig:ru_rsv_rdm}
\end{figure}

\section{Random state vector generator}

\label{sec:RSVG}

Pure states of $d$-dimensional quantum systems are described by unit
vectors in $\mathbb{C}^{d}$. The computational basis $|j\rangle=(\delta_{1j},\delta_{2j},\cdots,\delta_{dj})$
can be used to write any one of these vectors as
\begin{equation}
|\psi\rangle=\sum_{j=1}^{d}c_{j}|j\rangle,\label{eq:psi_cb}
\end{equation}
which are guaranteed to be normalized if $\sum_{j=1}^{d}|c_{j}|^{2}=1.$
A simple way to create random state vectors (RSVs) is by using normally
distributed real numbers to generate the real and imaginary parts
of the complex coefficients in Eq. (\ref{eq:psi_cb}), and afterwards
normalizing $|\psi\rangle$ (\texttt{opt\_rsvg = \textquotedbl{}gauss\textquotedbl{}}).

Using the polar form for the coefficients in Eq. (\ref{eq:psi_cb}),
$c_{j}=|c_{j}|\mathrm{e}^{i\phi_{j}}$, and noticing that $|c_{j}|^{2}$
is a probability distribution, we arrive at our \emph{standard} method
(\texttt{opt\_rsvg = \textquotedbl{}std\textquotedbl{}}) for generating
RSVs. We proceed then by defining $|c_{j}|^{2}\eqqcolon p_{j}$ and
writing
\begin{equation}
|\psi\rangle=\sum_{j=1}^{d}\sqrt{p_{j}}\mathrm{e}^{i\phi_{j}}|j\rangle.
\end{equation}
Then we utilize the RPVG to get $\vec{p}=(p_{1},\cdots,p_{d})$ and
the RNG to obtain the phases $(\phi_{1},\cdots,\phi_{d})$, with $\phi_{j}$
uniformly distributed in $[0,2\pi]$. Using these probabilities and
phases we generate a RSV. See examples in Fig. \ref{fig:ru_rsv_rdm}.
For these two first methods, when $d\gg1$, $\mbox{CC}(\mbox{RSVG})\approx\mathcal{R}(d)+\mathcal{O}(d^{2})$. 

In addition to these procedures, we have included yet another RSVG
using the first column of a RU (\texttt{opt\_rsvg = \textquotedbl{}ru\textquotedbl{}}):
\begin{equation}
|\psi\rangle=(U_{11},U_{21},\cdots,U_{d1}).
\end{equation}

\section{Random density matrix generator}

\label{sec:RDMG}

Our \emph{standard} method (\texttt{opt\_rdmg = \textquotedbl{}std\textquotedbl{}})
for random density matrix (RDM) generation (see e.g. Refs. \cite{Zyczkowski_vol,Maziero_rqs}),
starts from the eigen-decomposition
\begin{equation}
\rho=\sum_{j=1}^{d}r_{j}|r_{j}\rangle\langle r_{j}|
\end{equation}
and creates the eigenvalues $r_{j}$ and the eigenvectors $|r_{j}\rangle=U|j\rangle$
using, respectively, the RPVG and RUG described before. So, in this
case, $\mbox{CC}(\mbox{RDMG})\approx\mbox{CC}(\mbox{RPVG})+\mbox{CC}(\mbox{RUG})+\mathcal{O}(d^{6})\approx\mathcal{R}(d^{2})+\mathcal{O}(d^{6})$.

We can also produce RDMs by normalizing matrices with independent
complex entries normally distributed, named Wishart or Ginibre matrices
(\texttt{opt\_rdmg = \textquotedbl{}ginibre\textquotedbl{}}),
\begin{equation}
\rho=\frac{GG^{\dagger}}{||G||_{2}^{2}},
\end{equation}
where $||G||_{2}=\sqrt{\mathrm{Tr}(G^{\dagger}G)}$ is the Hilbert-Schmidt
norm \cite{Zyczkowski_induced,Zyczkowski_rrho}. A related method,
which produces RDMs with Bures measure (\texttt{opt\_rdmg = \textquotedbl{}bures\textquotedbl{}}),
uses
\begin{equation}
\rho=\frac{(\mathbb{I}+U)GG^{\dagger}(\mathbb{I}+U^{\dagger})}{||(\mathbb{I}+U)G||_{2}^{2}},
\end{equation}
with $U$ being a random unitary \cite{Zyczkowski_bures}. At last,
one can also generate RDMs via partial tracing a random state vector
$|\psi_{ab}\rangle$ \cite{Botero}:
\begin{equation}
\rho=\mathrm{Tr}_{b}(|\psi_{ab}\rangle\langle\psi_{ab}|);
\end{equation}
so \texttt{opt\_rdmg = \textquotedbl{}ptrace\textquotedbl{}}. See
examples in Fig. \ref{fig:ru_rsv_rdm}.

There are two issues arising from Fig. \ref{fig:ru_rsv_rdm} that
instantiate the utility of the numerical tool described in this article.
The first one regards quantum coherence quantification, which has
been rediscovered and formalized in the last few years \cite{Baumgratz,Winter_coh}.
We see that, while the average relative entropy of coherence concentrates
around a certain value, the $L_{1}$-norm coherence keeps growing
with the dimension $d$. Such kind of qualitative difference, promptly
identified in a simple numerical experiment, points towards a path
that can be taken in order to identify physically and/or operationally
relevant coherence quantifiers. The other issue refers to the too
fast concentration of measure reported in Ref. \cite{Maziero_rqs};
and which gains more physical appeal with the too entangled state
space reached by the last three RDMGs described in the last section.

It seems legitimate regarding the most random ensemble of quantum
states as being the one leading to minimal knowledge; which can, by
its turn, be identified with maximal symmetry \cite{Hall_rrho}. Thus,
for pure states we require such ensemble to be invariant under unitary
transformations (UTs), what implies in no preferential direction in
the Hilbert space. An ensemble of pure states drawn with probability
density invariant under UTs is said to be generated with Haar measure.
The same is the case for random unitaries \cite{Mezzadri}. We observe
that all random unitary generators and random state vector generators
described here produce Haar distributed random objects. 

In the general case of density matrices, invariance under UTs only
warrants ignorance about direction in the state space, but implies
nothing with respect to the eigenvalues distribution. In this regard,
in general, different metrics lead to distinct probability densities,
which are then used for constructing methods to create random density
matrices accordingly. Therefore, as advanced in Ref. \cite{Hall_rrho},
this situation calls for the application of physical or conceptual
motivations when choosing a RDMG. In this sense, we think that the
too fast concentration of measure issue, in conjunction with the well
known difficulty of preparing entangled states in the laboratory,
favors the standard random density matrix generator described above.

\section{Concluding Remarks}

\label{sec:tests}

To summarize, in this article we described Fortran codes for the generation
of random numbers, probability vectors, unitary matrices, and quantum
state vectors and density matrices. Our emphasis here was more on
ease of use than on sophistication of the code. For this is the starting
point for the development of a Fortran Library for Quantum Information
Science. In addition to including new capabilities for the generators
described here and to optimize the code, we expect to develop this
work in several directions in the future. Among the intended extensions
are the inclusion of entropy and distinguishability measures, non-classicality
and correlation quantifiers, simulation of quantum protocols, and
remote access to quantum random number generators. Besides, in order
to mitigate the explosive growth in complexity that we face in general
when dealing with quantum systems, $d=\dim\mathcal{H}\propto\exp(\mbox{No. of parties}),$
it would be fruitful to parallelize the code whenever possible.

We performed some simple tests and calculations for verification of
the code's basic functionalities. Some of the results are reported
in Figs. \ref{fig:rn_rpv} and \ref{fig:ru_rsv_rdm}. The code used
for these and other tests is also included and commented, but we shall
not explain it here. Several matrix functions are provided in the
files \texttt{matfun.f90} and \texttt{qnesses.f90}. For instructions
about how to compile and run the code see the readme file. In our
tests, we used BLAS 3.6.0, LAPACK 3.6.0 (see installation instructions
in \cite{inst_lapack}), and the GNU Fortran Compiler version 5.0.0.
A MacBook Air Processor 1.3 GHz Intel Core i5, with a 4 GB 1600 MHz
DDR3 Memory and Operating System OS X El Capitan Version 10.11.2 was
utilized. The code, and related files, can be downloaded in \cite{code_arXiv}
or \cite{code_GitHub}.
\begin{acknowledgments}
This work was supported by the Brazilian funding agencies: Conselho Nacional de Desenvolvimento Cient\'ifico e Tecnol\'ogico (CNPq), under processes 441875/2014-9 and 303496/2014-2, Instituto Nacional de Ci\^encia e Tecnologia de Informa\c{c}\~ao Qu\^antica (INCT-IQ), under process 2008/57856-6, and Coordena\c{c}\~ao de Desenvolvimento de Pessoal de N\'{i}vel Superior (CAPES), under process 6531/2014-08. I gratefully acknowledge the hospitality of the Physics Institute and Laser Spectroscopy Group at the Universidad de la Rep\'{u}blica, Uruguay, where this work was completed. I also thank Carlos Alberto Vaz de Moraes J\'unior for useful suggestions regarding the creation of Fortran libraries and one of the Reviewers by his/her constructive comments and suggestions.\end{acknowledgments}

\end{document}